\documentclass{article}
\usepackage{geometry}
\usepackage{booktabs}
\usepackage{amsmath,amssymb,amsthm,mathrsfs,amsfonts,dsfont} 
\usepackage[inline]{enumitem}
\theoremstyle{plain}
\newtheorem{assumption}{Assumption}
\newtheorem{remark}{Remark}
\newtheorem*{theorem*}{Theorem}

\newtheorem{theorem}{Theorem}
\newtheoremstyle{exampstyle} %https://en.wikibooks.org/wiki/LaTeX/Theorems
  {\topsep=0 pt} % Space above
  {\topsep=0 pt} % Space below
  {} % Body font
  {} % Indent amount
  {\bfseries} % Theorem head font
  {.} % Punctuation after theorem head
  {.5em} % Space after theorem head
  {} % Theorem head spec (can be left empty, meaning `normal')

\theoremstyle{exampstyle} \newtheorem{example}{Example}
\theoremstyle{exampstyle} \newtheorem{definition}{Definition}
\theoremstyle{exampstyle} 

\newtheorem*{notations*}{Notations}
\usepackage{tikz}
\usetikzlibrary{arrows, positioning}
\usetikzlibrary{shapes, arrows, decorations.pathmorphing}
\usepackage{siunitx} %Show units
\usepackage{algorithm, algorithmicx}
\usepackage{algpseudocode}
\DeclareMathOperator*{\argmax}{arg\,max}
\usepackage{graphicx}
\usepackage[subrefformat=parens,labelformat=parens]{subfig}
\title{Distributed, Private, and Derandomized Allocation Algorithm for EV Charging}
\author{
Hamid Nabati\footnote{Concordia University, Montreal, Quebec, Canada, (h\_nabati@encs.concordia.ca)}, 
Jia Yuan Yu\footnote{Concordia University, Montreal, Quebec, Canada, (jiayuan.yu@concordia.ca)} 
}
\begin{document}
\maketitle
%\title{Distributed, Private, and Derandomized Allocation Algorithm for EV Charging}  % put your title here!
% AAMAS: submissions are anonymous for most tracks
%\author{Paper \#70}  % put your paper number here!
\begin{abstract}  % put your abstract here!
  Efficient resource allocation is challenging when privacy of users
  is important. Distributed approaches have recently been
  used extensively to find a solution for such problems. In this work,
  the efficiency of distributed AIMD algorithm for allocation
  of subsidized goods is studied. First, a suitable
  utility function is assigned to each user describing the amount of satisfaction that it has from allocated resource. Then the resource allocation is defined as a
  \emph{total utilitarianism} problem that is an optimization problem
  of sum of users utility functions subjected to capacity
  constraint. Recently, a stochastic state-dependent variant of AIMD
  algorithm is used for allocation of common goods among users with
  strictly increasing and concave utility functions. Here, the stochastic AIMD algorithm is derandomized and
  its efficiency is compared with the stochastic version.  Moreover, the
  algorithm is improved to allocate subsidized goods to users with concave and
  nonmonotonous utility functions as well as users with Sigmoidal
  utility functions.  
  To illustrate the effectiveness of the proposed
  solutions, simulation results is presented for a public
  renewable-energy powered charging station in which the electric
  vehicles (EV) compete to be recharged.
\end{abstract}
%\keywords{Distributed Resource Allocation; Total Utilitarianism; AIMD
%Algorithm; Electric Vehicle (EV) Charging}  % put your semicolon-separated keywords here!
\section{Introduction}\label{introduction}
In many real-world applications, the goal is allocating scarce
resources among $n$ users in order to achieve maximum total utility so-called
\emph{total utilitarianism}. 
The concept of the \emph{utility} here represents the satisfaction level of each user from the allocated resources. It normally leads to solve an optimization problem that the objective function is the sum of users utility functions subjected to capacity and other constraints. Mathematically speaking, we have
\begin{equation} \label{Eq:NUM}
  \begin{aligned}
    & \underset{x_1, \dots, x_n}{\text{maximize}}
    & & \sum_{i=1}^{n}{u_i(x_i)} \\
    & \text{subject to}
    & & \sum_{i=1}^{n}{x_i\leq C} \, , \\
    &&& x_i\geq0 \; , \; i=1, \dots, n \, ,
  \end{aligned}
\end{equation}
where $u_i$ and $x_i$ denote each user $i$'s utility function and allocated resource, respectively and $C>0$ denotes the capacity constraint. Each user's utility function $u_i$ and number of user involved in resource allocation problem are unknown and reaching to capacity constraint $\sum_{i=1}^{n}{x_i\leq C}$ is just informed by a notification. \\
To solve the optimization problem given by~\eqref{Eq:NUM}, there are two main solution approaches:
\emph{centralized} and \emph{distributed}. Centralized solutions are
more efficient since users first admit their individual utility
functions to a decision maker, which then solves the optimization
problem to find the optimal allocated resources. However, users
utility functions are private information and the drawback is that
users' privacy protections is challenged.
Distributed allocation is a key concept to resolve this conundrum,
\emph{i.e.}, to efficiently allocate resources while preserving
privacy. In distributed resource allocation, a set of users must
autonomously assign their resources with respect to certain criteria
and the main goal is to reach the \emph{global optimum}.\\
To model different problems, we need to approximate each user's satisfaction with a suitable utility function. We consider three following cases.  First, for allocation of \emph{common goods}, where users do not pay a fee per use, we adopt \emph{concave and strictly increasing} utility functions, since it provides mathematical tractability~\cite{boyd2004convex} however limits its applicability. 
Second, for allocation of subsidized goods, where the fee per use is shared with the entire population, each user payoff function is defined as the difference between the user utility function and the cost of received resources. Therefore, considering concave and strictly increasing utility functions, users payoff functions are concave and nonmonotonous.
Third, consider allocation of goods that are only useful in sufficient quantities. Herein, each user's satisfaction ideally described by a discontinuous Step function that we approximate with a continuous Sigmoidal utility function~\cite{abdel2014utility}.
\begin{example} [EV Charging]
Recent studies reveal that a fuel-driven vehicle can produce less greenhouse gas emissions than an EV if the recharging energy is entirely produced
  by coal-fired power plants. Therefore, local stations for charging
  EVs from renewable energy significantly contributes to achieve real
  environmental benefits~\cite{xydas2016multi}.
Imagine a charging station of Electric Vehicles (EVs) whose power
  supplies from renewable energy (e.g., solar, wind). Such
  stations have limited available resources and demand for these
  finite amounts of energy is also increasing.  The users are EV owners who connect
  their vehicles to the charging station, and intuitively some of users need their vehicles more than others (e.g., the handicapped, elderly, and parents with young kids to pick-up after work).
   A private utility function
  determines the level of satisfaction for each EV owner whose EV is
  connected to the station to be charged.  As the demand for the
  resource overwhelms the capacity, every individual who consumes an
  additional unit directly harms others who can no longer enjoy the
  benefits. However, since the return of EVs to charging station is
  non-deterministic, it seems reasonable to assume that EV owners are
  greedy and prefer to charge their own EVs regardless of others due
  to avoid range anxiety. The users' utility functions are chosen normalized logarithmic to represent strictly increasing concave functions as well as Sigmoidal to approximate Step functions.
   \begin{figure}[h]
  \centering
  \includegraphics{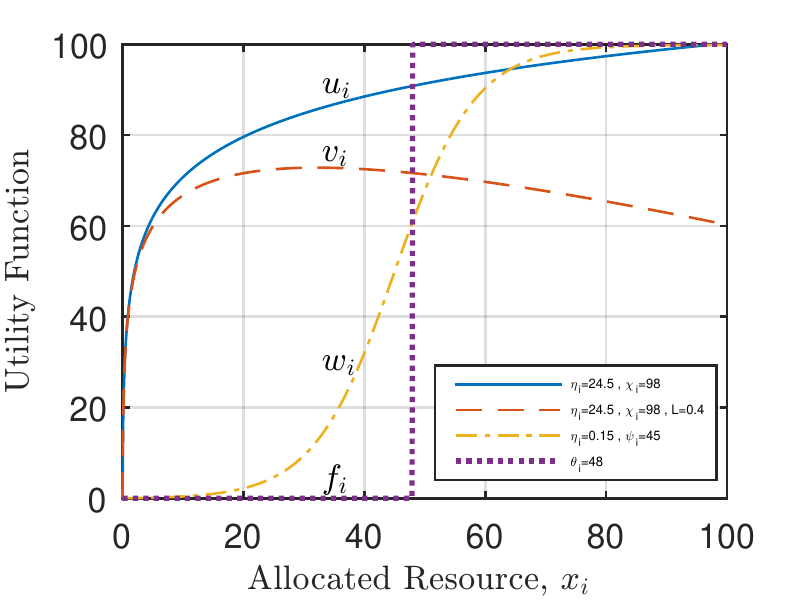}
  \caption{Utility functions: normalized logarithmic (strictly-increasing concave)
    utility function $u_i$ and corresponding nonmonotonous
  payoff functions $v_i$ when $L=0.4$, as well as  a discontinuous Step utility
    function~$f_i$ and corresponding approximate continuous Sigmoidal utility
    function~$w_i\,$. }\label{fig:uvsw}
\end{figure}
\end{example}
\begin{figure*}
  \begin{center}
    \begin{minipage}{.4\textwidth}
      \begin{tikzpicture}[>=latex']
 	\node (x) at ( 1.25,-2) [rectangle,draw,fill=blue!20,minimum
        width=6mm ,minimum height=45mm] {}; \node (x1) at ( 1.25,0)
        [rectangle,minimum size=6mm] {}; \node (input1) at
        (0,0){$u_1$}; \draw[->] (input1) -- +(x1);
	\node (x3) at ( 0,-1) [] {\vdots}; \node (xi) at ( 1.25,-2)
        [rectangle,minimum size=6mm] {}; \node (inputi) at
        (0,-2){$u_i$}; \draw[->] (inputi) -- +(xi); \node (x5) at (
        0,-3) [] {\vdots}; \node (xn) at ( 1.25,-4) [rectangle,minimum
        size=6mm] {}; \node (inputn) at (0,-4){$u_n$}; \draw[->]
        (inputn) -- +(xn); \node (n) at ( 0,1) [] {$n$}; \draw[->] (n)
        -- +(x1); \node (C) at ( 1,1) [] {$C$}; \draw[->] (C) --
        +(x1); \node (x*) at (3.25,-2)[]{$x_1^*, \dots, x_n^*$};
        \draw[->] (x) -- +(x*);
      \end{tikzpicture} \caption*{(a)}
    \end{minipage} %\hspace{1mm}%
    \begin{minipage}{.4\textwidth}
      \begin{tikzpicture}[>=latex', every node/.style={transform
          shape}]
	\node (empty) at (0,1){}; \node (input1) at (-0.3,0){$u_1$};
        \node (x1) at ( 1,0) [rectangle,draw,fill=blue!20,minimum
        size=6mm] {}; \draw[->] (input1) -- +(x1); \node (x1t) at
        (3.75,0){$\ x_1{(t)}, \dots\ \xrightarrow{t\rightarrow \infty}
          x_1^*$}; \draw[->] (x1) -- +(x1t);
	\node (x3) at (1.2,-1) [] {\vdots}; \node (xi) at ( 1,-2)
        [rectangle,draw,fill=blue!20,minimum size=6mm] {}; \node
        (inputi) at (-0.3,-2){$u_i$}; \draw[->] (inputi) -- +(xi);
        \node (xit) at
        (3.75,-2){$\ x_i{(t)}, \dots\ \xrightarrow{t\rightarrow
            \infty} x_i^*$};
        \draw[->] (xi) -- +(xit); \node (x5) at ( 1.2,-3) [] {\vdots};
        \node (xn) at ( 1,-4) [rectangle,draw,fill=blue!20,minimum
        size=6mm] {}; \node (inputn) at (-0.3,-4){$u_n$}; \draw[->]
        (inputn) -- +(xn); \node (xnt) at
        (3.75,-4){$\ x_n{(t)}, \dots\ \xrightarrow{t\rightarrow
            \infty} x_n^*$};
        \draw[->] (xn) -- +(xnt); \node (plus) at (4.2,1) []
        {$\bigoplus$}; \node (feedback) at (2,1) [rectangle,draw]
        {$\sum_{i=1}^{n}{x_i{(t)}} > C $}; \node (delay) at (-0.3,1)
        [rectangle,draw] {$\vartriangle $}; \node (signal) at (-1.8,1)
        [] {$s{(t-1)}$}; \draw[->,dashed] (feedback) -- +(delay);
        \draw[->,dashed] (delay) -- +(signal); \draw[->,dashed] (plus)
        -- +(feedback); \node (fake1) at (3,0) [] {}; \node (fakei) at
        (3,-2) [] {}; \node (faken) at (3,-4) [] {}; \draw[->,dashed]
        (fake1) -- +(plus); \draw[->,dashed] (fakei) -- +(plus);
        \draw[->,dashed] (faken) -- +(plus); \draw[->,dashed] (signal)
        -- (x1.west);
        \draw[->,dashed] (signal) -- (xi.west); \draw[->,dashed]
        (signal) -- (xn.west);
      \end{tikzpicture} \caption*{(b)}
    \end{minipage}%
  \end{center}
  \caption{Resource allocation solution approaches, (a) centralized ,
    (b) distributed iterative.}\label{BlockDiagram}
\end{figure*}
In this paper, we propose a \emph{distributed and iterative} algorithm, that is a computationally efficient and
private solution of the resource allocation problems.
A stochastic version of the algorithm was used
for common goods such as clean air and access to public
road in~\cite{wirth2014nonhomogeneous}.
We extend the applications
and we make the following specific contributions. We propose derandomized version of AIMD algorithm
  to allocate common goods to users with strictly increasing, concave utility functions. We also propose AIMD algorithm to allocate subsidized
  goods, where the fee per use is shared with the entire population, to users with concave and nonmonotonous utility functions.  We extend the results to propose a variant of AIMD
  algorithm to allocate common goods to users with Sigmoidal
  utility functions.\\
The rest of this paper is organized as follows. Section~\ref{sec:DRA} presents the problem formulation. In section~\ref{sec:AIMDAlg}, we propose variants of AIMD distributed algorithm for
allocation of subsidized (and common) goods among users based on
their specific utility functions. Section~\ref{sec:SimulationAIMD}, includes EV charging simulation setup and discuss the numerical results. Section~\ref{sec:CFW} concludes the paper.
\section{Distributed Resource Allocation } \label{sec:DRA} 
In this section, we formally define resource allocation problem using utility
function concept for both centralized and distributed solution
approaches. The objective is to determine each
user's optimal allocated resource at which maximum total utility is
achieved.
\begin{remark}
  In real life applications, the resource could be time-slotted like
  energy in \si{kWh} or time-varying like power in \si{kW}. Although,
  we consider a time-slotted optimization problem, we can use the proposed solution for any time-varying situations without changing the results.
\end{remark}
%\begin{notations*}
%Boldface letters (e.g. $\textbf{x}$) stand for vectors, $x_i$ refers to the $i$-th element of vector $\textbf{x}$ and $x_i(t)$ refers to $t$-th iteration of $x_i$. The notation $\mathds{1}_{[\text{Boolean expression}]}$ denote the indicator function that is equal $1$ if expression is true and $0$ otherwise.
%\end{notations*}
\subsection{Baseline and Problem Formulation}\label{Sec:BPFO} %Baseline
Consider $n$ users utilize a shared limited resource $C>0$, and let $x_i\geq0$ represents the possible allocated resource to each user $i=1, \dots,  n$.
% A column vector $\textbf{x}=[x_1 \, \dots \, x_n]^\top$ with $n$ real elements denoted by $\mathbb{R}^n$, where $x^\top$ indicates the transpose of $x$.
We attribute a \textit{utility},
\textit{i.e.}, a measure of satisfaction, to each user $i$ who takes
advantage of the common resource and describe it by means of a
\textit{utility function}. The utility function
$u_i:\mathbb{R}_+ \rightarrow \mathbb{R}_+ $, assigns a non-negative
real number to each possible value of allocated resource $x_i$, to
represent the level of satisfaction for each user $i$ or quality of
service (QoS).\\
A class of centralized resource allocation
problems can be formulated as a nonlinear continuous
optimization problem~\eqref{Eq:NUM} that is also represented in 
Figure \ref{BlockDiagram}a.
In such problems, a central decision maker calculates the
optimal solution vector $\textbf{x}^*=[x_1^*, ..., x_n^*]^\top$, by collecting all information
regarding each user's utility function $u_i$, capacity constraint $C$,
and number of users $n$. 
Although centralized solution approaches focus to determine efficient
resource allocation, in many realistic applications, it is neither
applicable nor desirable~\cite{marden2013distributed} since it
violates users' privacy.\\
Figure~\ref{BlockDiagram}(b) depicts a class
of \emph{distributed (and iterative)} approach to resource allocation
problems in which allocations emerge as the result of an iterative of
local procedures. In other words, a set of users locally make
decisions regarding their resources autonomously. To this end, an
algorithm is used to assign each user $i$ an allocated resource
$x_i{(t)}$ in time steps (iterations) ${1, \dots, t}$. In each
iteration, each user's algorithm update user's allocated resource
$x_i{(t)}$ locally by choosing one of these options: increase,
decrease or no-change compared with previous iteration
$x_i{(t-1)}$. The increase option continues until receiving one bit
signal $s{(t-1)}$, that notify capacity constraint
$\sum_{i=1}^{n}{x_i{(t)}}> C$ is violated and algorithm, based on a
certain probability, choose one of the following options: decrease or
no-change. When the capacity is available again
$\sum_{i=1}^{n}{x_i{(t)}} \leq C$, the increase option of the
algorithm restarts immediately. The procedure repeats until the number
of iterations is large enough $t$, and users' allocated resource
converge to the optimal allocation $x^*$.\\
In order to quantify efficiency of distributed resource allocation,
we express the efficiency as follows:
\begin{equation} \label{Eq:efficiency}
  \text{efficiency at time }  t = \frac{ \lim_{t\rightarrow\infty}  \sum_{i=1}^{n}{u_i(x_i(t))}}{\sum_{i=1}^{n}{u_i(x_i^*)}} \, , \\
\end{equation}
where the numerator is the output of proposed distributed
algorithm and the denominator is the solution of an interior-point optimization algorithm.
In the following sections, we show that the efficiency converges to $1$.
\section{AIMD Algorithm}\label{sec:AIMDAlg}
In this section, we describe several variants of AIMD algorithm.
\begin{remark}
AIMD \emph{(Additive Increase Multiplicative Decrease)} is a distributed and iterative algorithm that is used widely to control congestion in computer networks. The objective of AIMD is to
  determine the share of the resource for each user while total
  demands remains less than the available capacity, and where the
  limited communication in the network is desired. The AIMD algorithm, in its basic version,
  is composed of two procedures. In the additive increase (AI) phase,
  users continuously request for more available resource of the
  network until receiving a notification that the aggregate amount of
  available resource has been exceeded. Then the multiplicative
  decrease (MD) phase occurs and users respond to the notification by
  reducing their share proportionally. The AI phase of the algorithm
  restarts again immediately and this pattern is repeated by each
  active user in the network~\cite{corless2016aimd}.
\end{remark}
%\subsection{Common Goods}
For the sake of representation, we reproduce the stochastic allocated-dependent version of AIMD algorithm that is used for allocation of common goods among users with concave and strictly increasing utility functions. 
We shall not describe the guaranties of the algorithm here, rather we refer the interested readers to~\cite{wirth2014nonhomogeneous} for details.
Thus, we consider a further and substantial
assumption, so-called \emph{concavity assumption}, for users
utility functions which provide mathematical tractability of
optimization problem \eqref{Eq:NUM}.
\begin{assumption}[]\label{as:uconincdif} (Concavity Assumption) The
  utility functions $u_i:\mathbb{R}_+ \rightarrow \mathbb{R}_+$,
  \begin{enumerate*}[label=(\roman*)]
  \item are strictly increasing functions of $x_i$ with $u_i(0)=0$,
  \item are concave and continuously diffrentiable with domain
    $x_i\geq0$,  
  \end{enumerate*}
  where $x_i$ is the amount of resources allocated to user $i$.
\end{assumption}
\begin{example}
  We use normalized  logarithmic function~Equation~\eqref{Eq:NLU}
  as strictly increasing concave utility function that satisfies
  Concavity Assumption~\ref{as:uconincdif} in order to model the level
  of satisfaction of EV owners whose car is connected to the charging
  station to be charged.
    \begin{equation} \label{Eq:NLU} u_i(x_i)=100\frac{\log(1+\eta_i
      x_i)}{\log(1+\eta_i \chi_i)} \, ,
  \end{equation}
  where $\chi_i$ in \si{kWh} is the amount of allocated resource (EV
  charging) that gives $100$ unit utility to the user $i$. It also satisfies $u_i(0) = 0$.
  The parameter $\eta_i$ indicates how the charge
  needed urgently by effecting on the rate of utility percentage that
  is a function of allocated resource $x_i$. Intuitively higher values
  of $\eta_i$ yield higher utility to user $i$. Figure~\ref{fig:uvsw}
  represents a normalized logarithmic utility functions $u_i$ with
  $\eta_i=24.5,~\chi_i=98$.
  \end{example}
In AI phase, each active user $i$ continues to update its
allocated resource $x_i{(t)}$ upward by adding an amount of
\emph{growth factor} $\alpha \in (0, C)$ to its previous allocated
resource $x_i{(t-1)}$ while $\sum_{i=1}^{n}{x_i\leq C}$. When the
capacity limit has been violated, \emph{i.e.},
$\sum_{i=1}^{n}{x_i> C}$, users are notified to execute MD phase. Each
user will respond to the capacity signal independently with a certain
probability $\lambda_i\,$, by multiplying the previous allocated
resource $x_i{(t-1)}$ to a \emph{drop factor} $\beta \in (0,1)$ to
form current allocated resource $x_i{(t)}$.
The probability $\lambda_i$ at $t$-th iteration, for each user $i$,
depends on the long-term average allocated resource $\bar{x}_i{(t)} = \frac{1}{t+1} \sum_{k=1}^{t}{x_k} $~through the relation
$\lambda_i(\bar{x}_i{(t)})=\Gamma\frac{u'_i(\bar{x}_i{(t)})}{\bar{x}_i{(t)}}$,
where the parameter $\Gamma$ is chosen to ensure that
$0<\lambda_i(\bar{x}_i)<1$.
\begin{algorithm}
  \caption{AIMD for user $i$}
  \label{Al:AIMD}
  \begin{algorithmic}[1]
    \State Initialize $x_i{(0)}$ arbitrary \State Broadcast the
    parameter $\Gamma$ \For{time steps $t = 1, 2, 3, \dots$}
    \If{$\sum_{j=1}^{n}{x_j{(t)}}<C$} \State \hspace{-2mm}
    $x_i{(t+1)}=x_i{(t)}+\alpha$; \Else \State \hspace{-2mm}
    $x_i{(t+1)}=\beta x_i{(t)} $ with probability
    $\lambda_i(\bar{x}_i{(t)})=\Gamma\frac{u'_i(\bar{x}_i{(t)})}{\bar{x}_i{(t)}}$
    \State \hspace{-2mm} $x_i{(t+1)}=x_i{(t)}$ otherwise;
    \EndIf
    \EndFor
  \end{algorithmic}
\end{algorithm}
\subsection{Derandomized Algorithm}
%\textbf{On the first extension, the derandomization doesn't appear to have significant effects empirically, where DAIMD performs as well as AIMD. As such, I don't view this extension is particularly significant from an empirical standpoint. From an algorithmic standpoint, my impression is that stochastic algorithms typically are more robust than deterministic one. Unfortunately, the authors did not make a case on *why* derandomization is necessary or good. What does that buy us?}
The proposed stochastic AIMD Algorithm~\ref{Al:AIMD} is considered to have access to sources of \emph{perfect randomness}, \emph{ i.e.} unbiased and completely independent random variables, however in real-world implementation, the physical sources of randomness to which we have access may contain biases and correlations~\cite{vadhan2012}. Thus, the probabilistic method can also yield insight into how to
construct deterministic
algorithms~\cite{mitzenmacher2005probability}. We now propose a
variant of \emph{deterministic} AIMD for the same purpose. 
\begin{algorithm}
  \caption{DAIMD for user $i$}
  \label{Al:DAIMD}
  \begin{algorithmic}[1]
    \State Initialize $x_i{(0)}$ arbitrary \State Broadcast the
    parameter $\Gamma$ \For{time steps $t = 1, 2, 3, \dots$}
    \If{$\sum_{j=1}^{n}{x_j{(t)}}<C$} \State
    $x_i{(t+1)}=x_i{(t)}+\alpha$; \Else \State
    $\lambda_i(\bar{x}_i{(t)})=\Gamma\frac{u'_i(\bar{x}_i{(t)})}{\bar{x}_i{(t)}}$, \State
    $x_i{(t+1)}=\beta (1-\lambda_i) x_i{(t)} + \lambda_i x_i{(t)}$;
    \EndIf
    \EndFor
  \end{algorithmic}
\end{algorithm}
We show
that the strong convergence of derivative of utility function of
long-term average allocated resource $u'_i(\bar{x}_i{(t)})$, can be
used to allocate resource optimally. Therefore, there exists a derandomized algorithm with the same performance as the randomized algorithm.\\
We define a vector $z(t)$ consists of $d \in \mathbb{N}$ real-valued elements by
\begin{equation}
z(t) := [x_1 \, \dots \, x_n \  \bar{x}_1 \  \dots \, \bar{x}_n \  \lambda_1 \, \dots \, \lambda_n]^\top ,
\end{equation}
and let a class $\{f^{(t)}:\mathbb{R}^d \rightarrow \mathbb{R}^d, t=1,2, \dots\}$ of functions  such that
\begin{equation}
z(t) = f^{(t)}(z(t-1)).
\end{equation}
Each function $f_\ell^{(t)}:\mathbb{R}^d \rightarrow \mathbb{R}$, $\ell=1, \dots, d$ is also considered  ...\\
The allocated resource $x_i(t)$ at $t$-th iteration, for each user $i$, is calculated  as Equations~\eqref{eq:x_i(t)}.
\begin{equation}\label{eq:x_i(t)}
\begin{split}
x_i(t) = (x_i(t-1)+\alpha) \mathds{1}_{[\sum_{j=1}^{n}{x_j{(t)}}<C]} +(\beta (1-\lambda_i) x_i{(t)} \\
 + \lambda_i x_i{(t)}) \mathds{1}_{[\sum_{j=1}^{n}{x_j{(t)}} \geq C]}
\end{split}
\end{equation}

\begin{equation}
\bar{x}_i{(t)} = \frac{1}{t+1} \sum_{k=1}^{t}{x_k}
\end{equation}
Therefore, iterated function systems (IFS) is defined as:
We define
$\lambda_i(\bar{x}_i{(t)})=\Gamma\frac{u'_i(\bar{x}_i{(t)})}{\bar{x}_i{(t)}}$
and we use it in MD phase of the algorithm by
$x_i{(t+1)}=\beta (1-\lambda_i) x_i{(t)} + \lambda_i x_i{(t)}$ to
build DAIMD Algorithm~\ref{Al:DAIMD}.
\subsection{Subsidized Goods}
We extend resource allocation problem to subsidized goods where the
fee per use is shared with the entire population. Suppose if each user
$i$ is charged a constant price $L$ per unit of the received resources
$x_i$. Each user payoff function,
$v_i:\mathbb{R}_+ \rightarrow \mathbb{R}_+$ is defined as utility
function minus the cost of received resource as follows:
\begin{equation}\label{Eq:v_i}
  v_i(x_i)=u_i(x_i)-Lx_i \, .
\end{equation}
Recall each user utility function $u_i(x_i)$ is considered under
Concavity Assumption~\ref{as:uconincdif}, therefore, each user payoff
function \eqref{Eq:v_i} is a concave function but it is not
necessarily increasing. The centralized resource allocation problem
can then be formulated as follows:
\begin{equation} \label{Eq:NVM}
  \begin{aligned}
    & \underset{x_1, \dots, x_n}{\text{maximize}}
    & & \sum_{i=1}^{n}{v_i(x_i)} \\
    & \text{subject to}
    & & \sum_{i=1}^{n}{x_i\leq C} \, , \\
    &&& x_i\geq0 \; , \; i=1, \dots, n \, .
  \end{aligned}
\end{equation}
The optimization problem~\ref{Eq:NVM} in which the objective function
is non-negative sum of concave functions, is concave and there exists a
global optimal solution~\cite{boyd2004convex}.\par
\begin{example}
  Suppose normalized logarithmic function
  Equation~\eqref{Eq:NLU} to show each EV owner's satisfaction from receiving an amount of energy allocation $x_i\,$. The cost of allocated energy is defined as the price of energy $L$ in
monetary units, multiply in energy allocation $x_i\,$. Therefore, the payoff function is defined as follows:
  \begin{equation} \label{Eq:NLV} v_i(x_i)=100\frac{\log(1+\eta_i
      x_i)}{\log(1+\eta_i \chi_i)} -Lx_i \, .
  \end{equation}
  Figure~\ref{fig:uvsw} represents a normalized logarithmic
  utility function with $\eta_i=24.5~,~\chi_i=98$ compared with corresponding
  payoff functions $v_i$ when $L=0.4\,$.
\end{example}
We improve AIMD Algorithm~\ref{Al:AIMD} by controlling the allocation
do not exceed from maximum payoff of each user and design the PAIMD
Algorithm \ref{Al:PAIMD}. The control is applied locally since each
user $i$ calculates the optimal point
$x_i^*= \textstyle\argmax_{x_i \in \mathbb{R}_+} v_i(x_i), \
\forall i=1, \dots, n$
and then in each iteration, in the (AI) phase of the algorithm compare
it to allocated resource $x_i{(t)}+\alpha$ to choose the minimum
allocation.\\
\begin{algorithm}
  \caption{PAIMD for user $i$}
  \label{Al:PAIMD}
  \begin{algorithmic}[1]
    \State Initialize $x_i{(0)}$ arbitrary \State Each user $i$
    calculates
    $x_i^*= \displaystyle\argmax_{x_i \in \mathbb{R}_+} \ v_i(x_i), \
    \forall i=1, \dots, n$
    \State Broadcast the parameter $\Gamma$ \For{time steps
      $t = 1, 2, 3, \dots$} \If{$\sum_{j=1}^{n}{x_j{(t)}}<C$} \State
      \hspace{-2mm}
    $x_i{(t+1)} = \min( x_i^* \; , \; x_i{(t)}+\alpha)$ \Else \State
    \hspace{-2mm}
    $x_i{(t+1)}=\beta x_i{(t)} $ with probability
    $\lambda_i(\bar{x}_i{(t)})=\Gamma\frac{v'_i(\bar{x}_i{(t)})}{\bar{x}_i{(t)}}$
    \State \hspace{-2mm} $x_i{(t+1)}=x_i{(t)}$ otherwise;
    \EndIf
    \EndFor
  \end{algorithmic}
\end{algorithm}
\textbf{Based on $\chi_i and \eta_i$, define bounded value}\\
\subsection{Sigmoidal Utility Functions}
In this section, we model resource allocation problem~\eqref{Eq:NUM} using Sigmoidal users' utility functions that is defined as following:
\begin{definition}\label{Sigmoid_Utility_Function_Definition}
  The utility function of
  $w_i:\mathbb{R}_+ \rightarrow \mathbb{R}_+ $ is defined to be
  Sigmoidal if:
  \begin{enumerate*}[label=(\roman*)]
   \item The $w_i(0)=0$ and $w_i$ is strictly increasing function
    of $x_i$.
  \item $w_i(x_i)$ is continuously differentiable, with domain
    $x_i\geq0$.
  \item  $w_i(x_i)$ is convex for $x_i\leq \psi_i $ and is
    concave for $x_i \geq \psi_i$, which $\psi_i \in \mathbb{R}_+ $ is
    the inflection point.
  \end{enumerate*}
\end{definition}
\begin{algorithm}[]
  \caption{QAIMD for user $i$}
  \label{Al:QAIMD}
  \begin{algorithmic}[1]
    \State Initialize $x_i(0)$ arbitrary \State Broadcast the
    parameters $\Gamma_1$, $\Gamma_2$ \For{time steps
      $t = 1, 2, 3, \dots$} \If{$\bar{x}_i(t)<\psi_i$}
    \If{$\sum_{j=1}^{n}{x_j(t)}<C$} \State 
    $x_i(t+1)=\frac{1}{\beta} x_i(t)$ with probability
    $\lambda_i=\Gamma_1\frac{w'_i(\bar{x}_i(t))}{\bar{x}_i(t)}$
    \State $x_i(t+1)=x_i(t)$ otherwise; \Else \State 
    $x_i(t+1)= \max( 0 \; , \; x_i(t)-\alpha )$
    \EndIf
    \Else \State do AIMD with $\alpha$, ${\beta}$ and probability
    $\lambda_i=\Gamma_2\frac{w'_i(\bar{x}_i(t))}{\bar{x}_i(t)}$
    \EndIf
    \EndFor
  \end{algorithmic}
\end{algorithm}
\begin{example}
  (Why do we need Sigmoidal utility functions?). In some situations,
  such as charging an electric vehicle with the goal of reaching a
  predetermined destination (e.g., airport, home, etc.), the user
  receive negligible (or non) utility until a threshold of resource is
  reached (e.g., enough electric charge to arrive at the destination).
  Ideally, the best description of the utility
  function is through a discontinuous Step function as follows:
  \begin{equation} \label{Eq:Step} f_i(x_i)=\begin{cases}
      0 & \text{if $x_i < \theta_i$} \, ;\\
      100 & \text{if $x_i \geq \theta_i$} \, ,
    \end{cases}
  \end{equation}
  where $\theta_i$ shows the sufficient allocated resource that gives
  100 unit utility to user~$i$. \\
  Continious Sigmoidal utility functions may be used to approximate a
  step utility function to any arbitrary
  accuracy~\cite{udell2013maximizing}. Thus, we model EV owner's satisfaction with Sigmoidal utility function that is expressed by:
  \begin{equation} \label{Sigmoidal_Function} w_i(x_i)=
    \frac{100}{1+e^{-\eta_i(x_i-\psi_i)}}-\frac{100}{1+e^{\eta_i
        \psi_i}} \, ,
  \end{equation}
  where $\eta_i$ is the steepness of the curve that indicates how the
  charge is needed urgently for each user~$i$. The parameter $\psi_i$
  in \si{kWh} is the inflection point that achieving
  it satisfies the urgent need of user $i$. The function, also, 
  satisfies $w_i(0)=0$ and $\lim_{x_i\to+\infty} w(x_i)\approx100$.
  Figure~\ref{fig:uvsw} represents a Step utility function~$f_i$
  with $\theta_i = 48$ and an approximate corresponding Sigmoidal
  utility function~$w_i$ with~$\eta_i=0.15$ and $\psi_i=45\,$.
\end{example}
The QAIMD Algorithm~\ref{Al:QAIMD} represents the procedure of
allocation among users with Sigmoidal utility functions. The
key point is that in each iteration $(t)$, the long-term of allocated
resource $\bar{x}_i(t)$ is compared with each user $i$'s inflection
point $\psi_i$. If $\bar{x}_i(t)<\psi_i$, the increase phase is built
by multiplying the previous state $x_i(t)$ in a growth factor
$\frac{1}{\beta} >1$ to construct current state $x_i(t+1)$ with a
probability
$\lambda_i
(\bar{x}_i(t))=\Gamma_1\frac{w'_i(\bar{x}_i(t))}{\bar{x}_i(t)}$.
The decrease phase also is made by subtracting $\alpha$ from the
previous state. When $\bar{x}_i(t)\geq\psi_i$, the algorithm is work
with AIMD Algorithm~\ref{Al:AIMD} procedure. Note that there are two
parameters $\Gamma_1$, $\Gamma_2$ to ensure $0<\lambda_i(\bar{x}_i)<1$
in each case.
\section{Simulation} \label{sec:SimulationAIMD}
\begin{figure}[]
  \centering
    \vspace{-0.6cm}
  \subfloat[]{\includegraphics{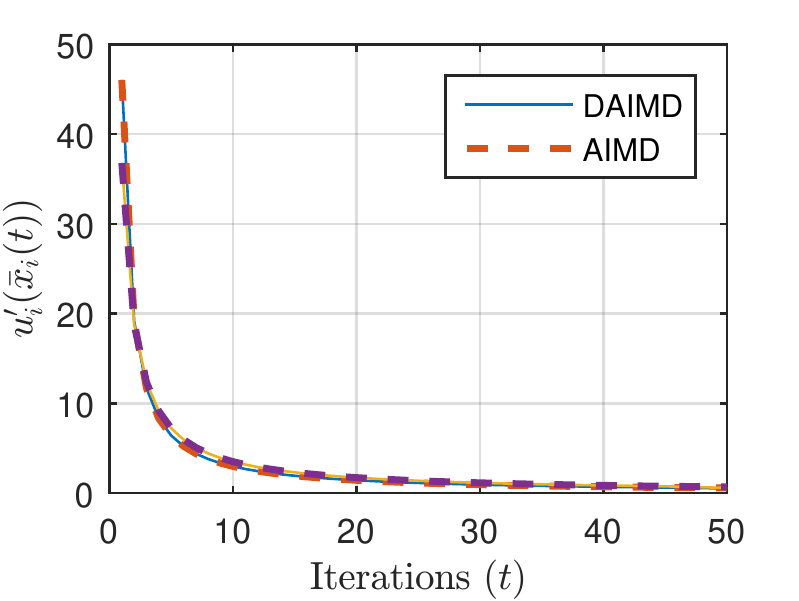} \label{fig:Duprime_xbar_t_compare}}
  \vspace{-0.4cm}
  \hspace{0.01 mm}
  \subfloat[]{\includegraphics{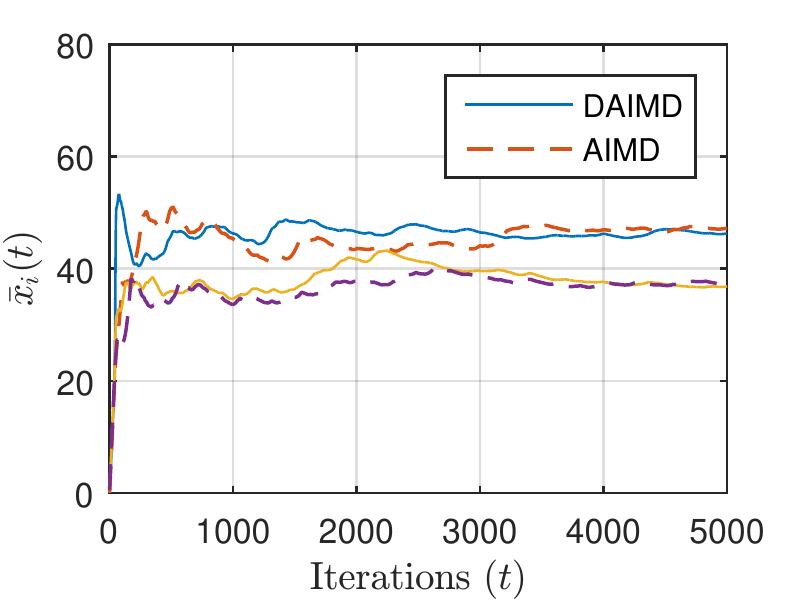}\label{fig:Dxbar_t_compare}}
    \vspace{-0.4cm}
    \hspace{0.01 mm}
  \subfloat[]{\includegraphics{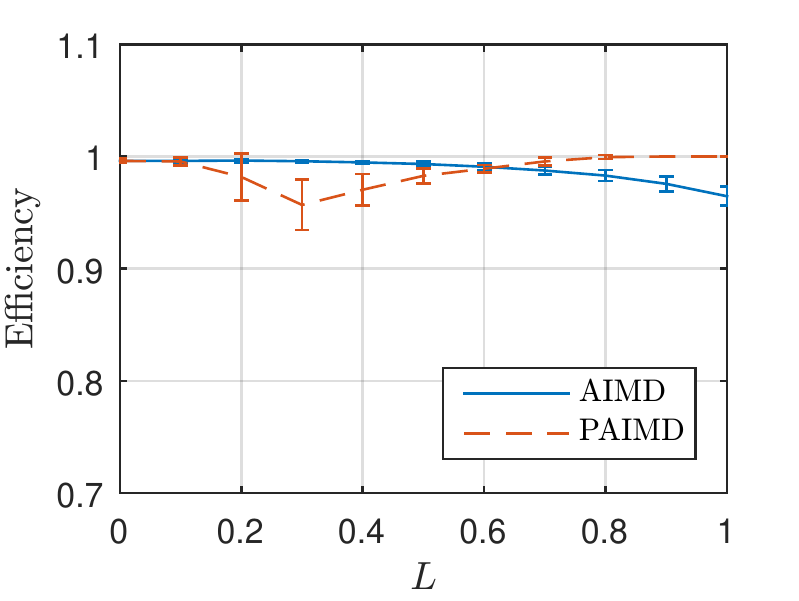} \label{fig:Efficiency}}
      \vspace{-0.4cm}
  \caption{ \protect\subref{fig:Duprime_xbar_t_compare} The deterministic
    derivative of payoff function $u'_i(\bar{x}_i{(t)})$ for two
    randomly selected users (solid lines) compared with corresponding
    stochastic ones (dashed lines),
    \protect\subref{fig:Dxbar_t_compare} the deterministic average of
    allocated resource $\bar{x}_i{(t)}$ to the stable point for two
    randomly selected users (solid lines) compared with corresponding
    stochastic ones (dashed lines), \protect\subref{fig:Efficiency} the efficiency of AIMD
    Algorithm~\ref{Al:AIMD} and PAIMD Algorithm~\ref{Al:PAIMD} for
    $L\in\{0, 0.1, \dots, 1\}$ calculated by Equation
    \eqref{Eq:efficiency}.}
\end{figure}
In order to figure out the effectiveness, the AIMD Algorithm~\ref{Al:DAIMD},~\ref{Al:PAIMD} and \ref{Al:QAIMD} was applied to various logarithmic and Sigmoidal utility functions with different parameters in MATLAB.\\
The resource allocation domain is considered as a charging
station whose power supplies from renewable energy (e.g. solar or
wind), with constant capacity  in \si{kWh} equal to $C=35\%$ of the sum of users utility functions when each
user receives $100$ unit satisfaction. The users $n=50$ are the EV
owners who connected their vehicles to the station for charging at the
same time.
%We define the utility $u_i$, of a greedy EV owner $i$, to
%be increasing to the charging status $x_i$ in \si{kWh} at which his EV
%is charged. In other words the EV owner's satisfaction is increasing
%to the extent that their EV is charged. 
The AIMD parameters are identical for all users with $\alpha=1$ and $\beta=0.85$. The parameter $\Gamma$ is also chosen to assure us the condition $\lambda_i(\bar{x}_i)\in (0,1)$ is
satisfied.\\
\begin{figure}[]
  \centering
  \vspace{-0.6 cm}
  \subfloat[]{\includegraphics{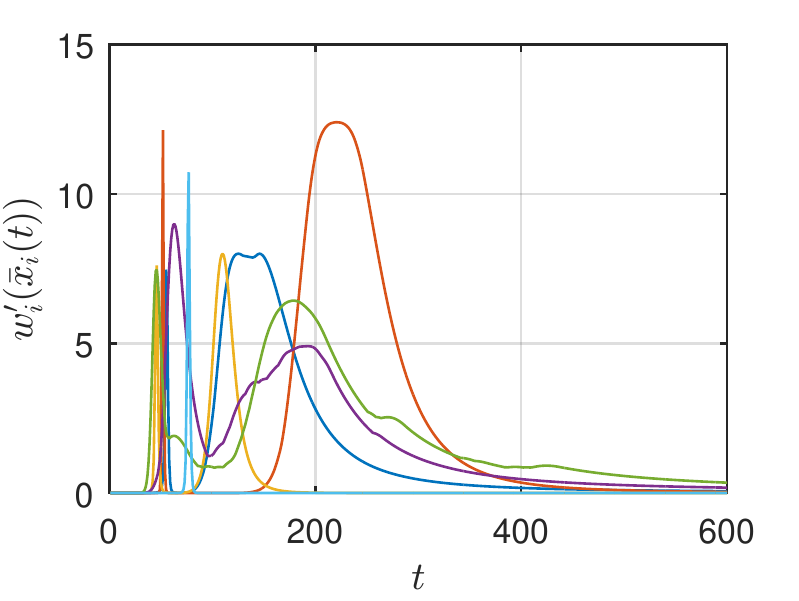} \label{fig:wprime_xbar_t}}
  \vspace{-0.4cm}
  \hspace{0.01 mm}
  \subfloat[]{\includegraphics{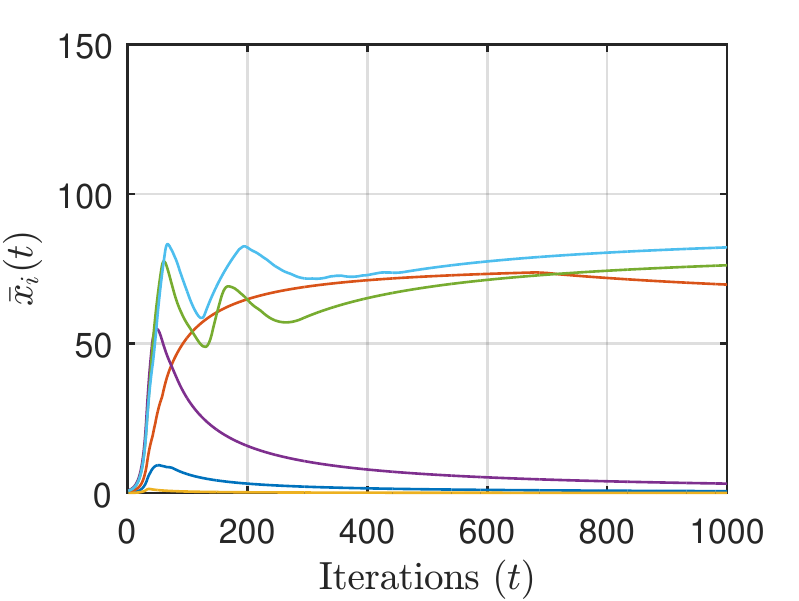} \label{fig:wxbar_t}}
  \vspace{-0.4cm}
  \hspace{0.01 mm}
  \subfloat[]{\includegraphics{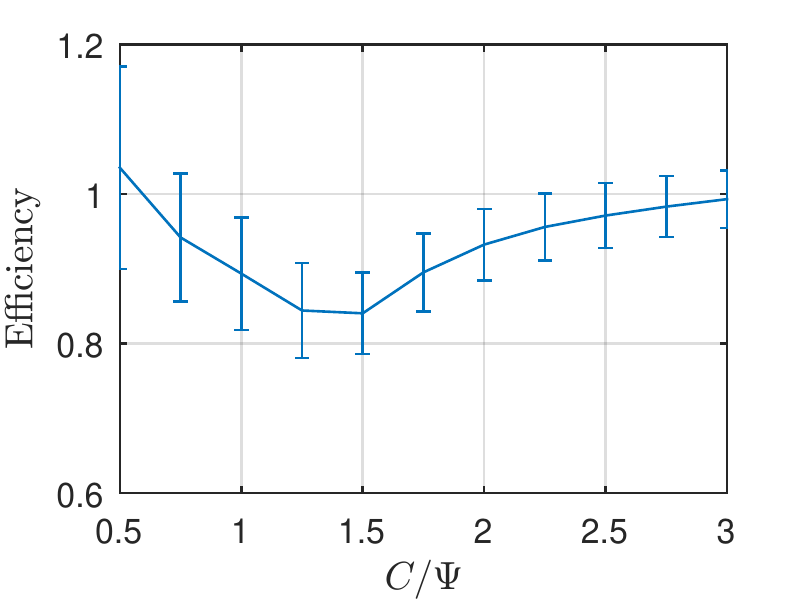} \label{fig:Result_Sigmoid}}
%  \vspace{-0.4cm}
  \caption{ \protect\subref{fig:wprime_xbar_t} The derivative of Sigmoidal
    utility functions $w'_i(\bar{x}_i{(t)})$ for six randomly selected
    users when $C / \Psi=1.5$, \protect\subref{fig:wxbar_t} the
    average of allocated resource $\bar{x}_i{(t)}$ to the stable
    point for six randomly selected users when $C / \Psi=1.5$,
    \protect\subref{fig:Efficiency} the efficiency of QAIMD
    Algorithm~\ref{Al:QAIMD} calculated by Equation
    \eqref{Eq:efficiency}.}
\end{figure}
%\subsection{Concave Utility Functions}
First, we adopt normalized logarithmic utility function expressed by~\eqref{Eq:NLU}, as a strictly increasing concave function
which satisfies Concavity Assumption~\ref{as:uconincdif}. We choose
$\chi_i$ independent uniformly distributed random number with support
$(40,60)$ and $\eta_i$ independent uniformly distributed random number
with support $(0,1)$.\\
We apply deterministic DAIMD~Algorithm~\ref{Al:DAIMD} for allocation
of power as a common good (no charging fee) to EVs connected to the charging
station. Figure~\ref{fig:Duprime_xbar_t_compare} shows a rapid convergence for derivative of payoff functions $u'_i(\bar{x}_i(t))$
when iteration $t$ increases. It also reveals the
coincidence of deterministic and stochastic versions of derivative of
payoff functions $u'_i(\bar{x}_i(t))$.
Figure~\ref{fig:Dxbar_t_compare} depicts the
value of long-term average state $\bar{x}_i(t)$ for two randomly
selected users and shows each of them converge to a stable value. It also represents that deterministic and stochastic version of average state
$\bar{x}_i(t)$ fluctuate differently but the long-term averages for
each user converge to optimal allocation. 
The efficiency of
deterministic DAIMD~Algorithm~\ref{Al:DAIMD}, calculated by
Equation~\eqref{Eq:efficiency}, in different runs are a real number in
the range of $(0.97, 0.99)$.\\
We also apply stochastic PAIMD~Algorithm~\ref{Al:PAIMD} for allocation of
power as a subsidized good to EVs connected to
charging station. Therefore, we consider the price per unit
$L \in \{0.1, 0.2, \dots, 1\}$ of the power $x_i$ in the
payoff function~\eqref{Eq:NLV}. Figure~\ref{fig:Efficiency}, depicts the efficiency of PAIMD Algorithm~\ref{Al:PAIMD} compared with AIMD Algorithm~\ref{Al:AIMD}, both calculated by Equation~\eqref{Eq:efficiency}. The efficiency of PAIMD~Algorithm~\ref{Al:PAIMD} decreases when the value of $L$ increases until receiving to a minimum value in ($L=0.3$). For large values of $L>0.3$ the efficiency of PAIMD Algorithm~\ref{Al:PAIMD} increases and converge to $1$.\\
%%%%%
Second, we model EV owner satisfaction with Sigmoidal utility function
that are expressed by Equation~\eqref{Sigmoidal_Function}. We choose
$\psi_i$ independent uniformly distributed random number with support
$(25,100)$ and $\eta_i$ independent uniformly distributed random
number with support $(0,25)$. The parameter $\Gamma_1$ and $\Gamma_2$ is also chosen to assure us the condition $\lambda_i(\bar{x}_i)\in (0,1)$ is
satisfied.
Figure~\ref{fig:wprime_xbar_t} depicts the derivative of utility
functions $w'_i(\bar{x}_i{(t)})$ for six randomly selected users. It
illustrates that the derivatives approach to zero as $t$ increase but
the convergence is slower than the derivatives of logarithmic utility
function $v'_i(\bar{x}_i{(t)}$ in Figure~\ref{fig:Duprime_xbar_t_compare} . In
Figure~\ref{fig:wxbar_t} the average of allocated resource
$\bar{x}_i{(t)}$ for six randomly selected users is displayed. It
shows $\bar{x}_i{(t)}$ approach to a constant value for some users and to $0$ for some other users.\\
Figure~\ref{fig:Result_Sigmoid} represents the efficiency of
QAIMD~\ref{Al:QAIMD}, calculated by Equation \eqref{Eq:efficiency},
for different capacity $C / \Psi=\{0.5, 0.75, \dots, 3\}$, where
$\Psi=\sum_{i=1}^{n} {\psi_i}$. For each user~$i$ the algorithm
decides between increasing allocated resource or decreasing it toward
zero. The efficiency of the algorithm is better for small values of
capacity constraint, but it decrease when capacity is around
$\Psi$. The efficiency improve again when $C/\Psi$ is large enough.

%\ifthenelse{\equal{\showsReport}{1}}{
%  \begin{Report}
\section{Loss of Efficiency Due to Competition} \label{sec:CSCR} 
In this section, we allow the individual users to act strategically as in a game.  
We consider a game in strategic form, where all users' utility functions are common knowledge. 
The resulting competition over a scarce
resource is reminiscent of the tragedy of the commons \cite{}. 
An user may deviate from the AIMD algorithm and strategically
request more resource in order to improve its
payoff.
Alternatively, an user may follow the AIMD algorithm but mispresent its utility function.
However, we show that, in some situations, the AIMD outcome and the game's Nash equilibrium are close to each other.
\subsection{Resource
Allocation as a Strategic Game} \label{Sec:Problem
  Setting as a Strategic Game} 
 Imagine a
resource allocation problem in which there are $n$ users, competing to
utilize a scarce fixed common resource of $C>0$. Each user $i$ chooses
his own consumption of resources $ x_i $ from a set of action space
$X_i = \{x_i \in \mathbb{R} \ | \ 0 \leq x_i \leq C \} $. A profile of
actions $x=(x_i, x_{-i})$ describe a particular combination of actions
chosen by all users, where $x_{-i}\in X_{-i}$ is a particular possible
of actions for all players who are not $i$.\par
Consuming an amount $x_i \geq 0$ gives user $i$ a benefit equal to
$u_i(x_i)$ when ${\sum_{j=1}^{n}{x_j}\leq C}$ and intuitively no other
users benefits from $i$'s choice. When $x_i$ increases or other users
consume more resources so that $\sum_{j=1}^{n}{x_j} > C$, the user get
nothing $u_i(x_i)=0$ because additional requested resources are not
provided. Then we define the payoff function $\tilde u_i(x_i, x_{-i})$
of a user $i$ from a profile of actions $x$ as
\begin{equation} \label{Eq:u_tilde} \tilde u_i(x_i,
  x_{-i})=\begin{cases}
    u_i(x_i) & \text{if $\sum_{j=1}^{n}{x_j}\leq C$} \, ;\\
    0 & \text{if $\sum_{j=1}^{n}{x_j}> C$} \, .
  \end{cases}
\end{equation}
Where the utility function $u_i(x_i)$ is considered to be concave,
strictly increasing, and continuously differentiable ,\textit{i.e.},
follows assumption \ref{as:uconincdif}.\par
The strategic game $(\mathcal{N},X_i,\tilde{u_i})_{i\in \mathcal{N}}$,
that have infinitely many pure strategies but utility functions are
not continuous, is discontinuous infinite strategic games. This
problem should be consider precisely because it may lead to problem of
nonexistence of unique Nash equilibrium.
\subsection{Nash Equilibrium}
To cut to the chase, the key notion to solve the \emph{strategic game}
$(\mathcal{N},X_i,\tilde{u_i})_{i\in \mathcal{N}}$, is the \emph{Nash
  equilibrium}, that is an outcome (a decision made by each player)
such that no player can improve his individual payoff through an
unilateral move. As stable situations, Nash equilibrium are often
considered to be the expected outcomes from interactions.
To solve for a Nash equilibrium we compute the best-response function
correspondence for each player and then find an action profile for
which all best-response functions are satisfied together.\par
To find a solution for Equation \eqref{Eq:u_tilde}, we first write out
each player $i$'s best-response correspondence and we consider that
given $x_{-i}$, player $i$ will want to choose an element in
$BR_i(x_{-i})$. Given $x_{-i}\in X_{-i}$ each player $i$'s best
response is the difference between $C$ and
$\sum_{j \neq i}^{n}{x_j} \,$. If user $i$ asks for more, then all
users get nothing while if asks for less then he is leaving some
resources unclaimed and therefore
\begin{equation}
  BR_i(x_{-i})= C-\sum_{j \neq i}^{n}{x_j} \, .
\end{equation}
It is easy to see from the best response correspondence that any
profile of demands $x_i \in [0, C]$ that add up to $C$ will be a Nash
equilibrium. Hence, each player $i$ is indifferent between all of his
requests $x_i \in [0, C]$ and the game is just not blessed with a
unique equilibrium and has an infinite number of equilibria. The
obvious problem with multiple equilibria is that the players may not
know which equilibrium will prevail. Hence, it is entirely possible
that a non-equilibrium outcome results because one player plays one
equilibrium strategy while a second player chooses a strategy
associated with another equilibrium~\cite{cachon2006game}.\par
It turns out that resource allocation encounters conflict over scare
resources that results from the tension between individual selfish
interests and common good. As Hardin stated in his article
\cite{hardin2009tragedy}, ``freedom in a commons brings ruin to all,''
that here means, social utility of an uncontrolled use of the common
resources that each user have the freedom to make choices, is worse
than if those choices were regulated. This results in the occurrence
of the phenomenon called \emph{tragedy of the commons}. In fact,
individual users acting independently according to their own
self-interest behave contrary to the common good of all users by
depleting that resource through their collective actions.\par
To solve this problem, we first need to bring back the continuity to
the payoff function Equation~\eqref{Eq:u_tilde}. Thus, we apply the
resource allocation back-off condition $\sum_{j=1}^{n}{x_j}> C$
directly to the payoff function for each user $i$. We define a concave
penalty function ${\tau:\mathbb{R}_+ \rightarrow \mathbb{R}_+} $ so
that $\tau(0)=1$ and $\tau(C)=0$ and multiply it to the payoff
function~\eqref{Eq:u_tilde}. To generalize, we also consider each unit
of resource costs $L$ and we have
\begin{equation} \label{Eq:v_tilde} 
  \begin{aligned}
& \tilde v_i(x_i, x_{-i})= u_i(x_i)  \, \tau \Big(\sum_{j=1}^{n}{x_j}\Big)-L x_i\, , \\
& \textrm{for all}  \quad x_i, x_j \in [0,\infty) \, .
  \end{aligned}
\end{equation}
\begin{example}
  Consider, for example, the concave penalty function $\tau(z)$ as
  follows:
  \begin{equation} \label{Eq:penalty} \tau \Big( z \Big) =
    \sqrt{1-\frac{z^p}{C^p} } \, \ ,
  \end{equation}
  where $z=\sum_{j=1}^{n}{x_j}$ and $p \in \mathbb{N}$. Intuitively,
  $\tau(0)=1$ and $\tau(C)=0$.\par
  Figure~\ref{fig:penalty} represents some examples of concave penalty
  functions Equation~\eqref{Eq:penalty} for
  $p\in\{1,2,4,8\}$. Although the larger values of $p$ reduce
  inefficiency of Nash equilibrium, however make calculations more
  complex. In realistic situation of EV charging, this function can be
  programmed to the charger and it works when the demand exceeds from
  capacity $C$. \par
  \begin{figure}[h]
    \centering
    \includegraphics{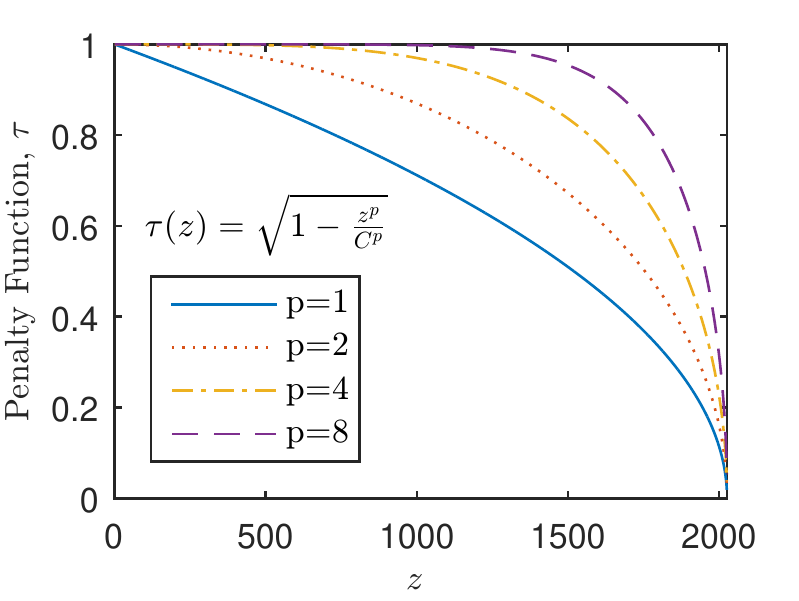}
    \caption{Penalty Function}\label{fig:penalty}
  \end{figure}
\end{example}
Since the payoff functions are continuous there is a strong result on
existence of the pure Nash equilibrium that is stated by
Theorem~\ref{Theorem:Debreu}~\cite{debreu1952social}.
\begin{theorem} \label{Theorem:Debreu} (Debreu, Glicksberg, Fan) An
  infinite strategic form game $\mathcal{G}=(\mathcal{N},X_i,f_i)_{i\in \mathcal{N}}$ such that for
  each $i \in \mathcal{G}$
  \begin{enumerate*}[label=(\roman*)]
  \item $X_i$ is compact and convex;
  \item $f_i(x_i,x_{-i})$ is continuous in $x_{-i}$;
  \item $f_i(x_i,x_{-i})$ is continuous and concave \footnote{in
      fact quasi-concavity suffices.} in $x_i$ .
  \end{enumerate*}
  % \vspace{-6mm}
  Then a pure strategy Nash equilibrium exists.
\end{theorem}
Another important question that arises in the analysis of strategic
form games is whether the Nash equilibrium is
unique. Theorem~\ref{Theorem:Rozen}, provides sufficient conditions
for uniqueness of an equilibrium in games with infinite strategy sets.
\begin{theorem}[Theorem
  1,~\cite{rosen1965existence}] \label{Theorem:Rozen} Consider a
  strategic form game
  $\mathcal{G}=(\mathcal{N},X_i,f_i)_{i\in \mathcal{N}}$ . For all
  $i \in \mathcal{G}$, assume that the action sets
  $X_i = \{x_i \in \mathbb{R}^{m_i} | h_i(x_i)\geq0\}$, where $h_i$ is
  a concave function, and there exists some
  $\tilde x \in \mathbb{R}^{m_i}$ such that $h_i(\tilde x_i)>0$
  . Assume also that the payoff functions $(f_i)_{i\in \mathcal{N}}$
  are diagonally strictly concave for $x \in X$ . Then the game has a
  unique pure strategy Nash equilibrium.  Where payoff functions
  $(f_i)_{i\in \mathcal{N}}$ are diagonally strictly concave for
  $x \in X$, if for every $x^{ne}, \bar{x} \in X$, we have
  $(\bar{x}-x^{ne})^\top \nabla f(x^{ne}) + (x^{ne}-(\bar{x})^\top
  \nabla f(\bar{x})>0$ .
\end{theorem}
The game $(\mathcal{N},X_i,\tilde{v}_i)_{i\in \mathcal{N}}$ has unique
Nash equilibrium that is calculated by maximizing user $i$'s payoff
function $\tilde v_i(x_i,x_{-i})$ and finding the solution to the
first order conditions. So, we write down the first-order condition of
user $i$'s payoff function as follows
\begin{equation} \label{Eq:First_Order} x_i^{ne} = \frac{\partial
    \tilde v_i(x_i,x_{-i})}{\partial x_i} =0 \, .
\end{equation}
We therefore have $n$ such equations, one for each player, and the
unique Nash equilibrium is the strategy profile $x^{ne}$ for which all
users in the network, the Equation \ref{Eq:First_Order} are satisfied
together, so that
\begin{equation} \label{Eq:xStar} x^{ne}=(x_i^{ne}, x_{-i}^{ne}), \
  x_i^{ne}=\displaystyle\argmax_{x_i \in [0,C]} \ \tilde
  v_i(x_i,x_{-i}), \ \forall i \in \mathcal{N} \, .
\end{equation}
When resource allocation problem form as a result of selfish
competition among users, the resulting stable solution may not, in
fact, be system optimal~\cite{lichter2011calculation}. In this
circumstance, we would like to measure inefficincy constituted due to
decentralized control. This is very important to decide whether a
decentralized mechanism can be applied, regarding the loss of
efficiency in comparison with the performance that would be obtained
with a central authority. \emph{Price of anarchy} (PoA)~
\cite{koutsoupias1999worst}, is a concept that quantifies this
inefficiency and is measured as the ratio between the worst
equilibrium and the centralized solution.
In the problem considered here, this notion will be slightly different
and defined as the efficiency of the unique Nash equilibrium of the
game $\mathcal{G}=(\mathcal{N},X_i,f_i)_{i\in \mathcal{N}}$ and the
optimal centralized solution of \eqref{Eq:NLV} as follows:
\begin{equation}
  PoA= \frac{ \sum_{i=1}^{n}{\tilde{v}_i(x_i^{ne})}}{\sum_{i=1}^{n}{v_i(x_i^*)}} \, ,\\
\end{equation}
where $x^{ne}$ is the unique Nash equilibrium given by
\eqref{Eq:xStar} and $x^*$ is the solution of \eqref{Eq:NLV} .
\subsection{Simulation} \label{Numerical Simulation, Competition} In
this section, we proceed to simulate resource allocation in
competition game to investigate in more details the inefficiency of
Nash equilibrium. For this purpose, suppose the EV charging station
settings of the section \ref{Sec:AIMD Numerical Simulation}. Each
user's utility function is considered as the normalized logarithmic
function \eqref{Eq:NLU} with uniformly distributed random parameters
of $\eta_i \in (0,25)$ and $\chi_i\in(25,100)$. We also consider the
concave penalty function Equation~\ref{Eq:penalty} with $p=1$ for
executing the simulation.  We start to simulate the problem for two
players. Consider the charging station with the limited resource of
$C=25$ \si{kWh} and two EV owners $i\in\{1 , 2\}$ which their EVs are
connected to the station for charging. Both players have normalized
logarithmic utility function Equation \eqref{Eq:NLU} with parameters
$\eta_1=15$, $\chi_1=30$ and $\eta_2=38$, $\chi_2=70$
respectively. Figure \ref{fig:contour_Csmall_tau} depicts inefficiency
of distributed competitive resource allocation in two-player game
,\emph{i.e.}, best response functions lines intersection, compared
with optimal solution $U(x_1^* , x_2^*)$.\par
\begin{figure}[H]
  \centering
  \includegraphics{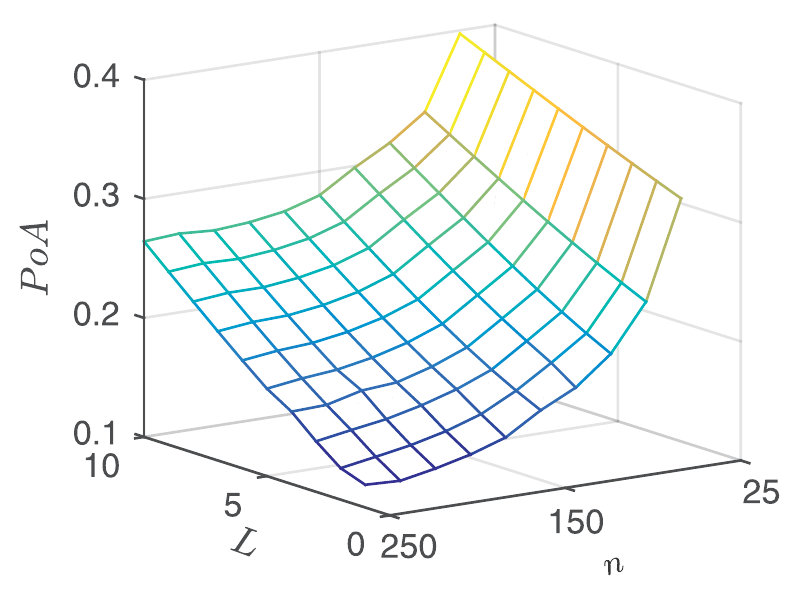}
  \caption{Price of Anarchy (PoA)}\label{fig:PoA}
\end{figure}

Now consider the same setting for a charging station with $n=50$
users. Figure plots social optimum of Nash
equilibrium $ \sum_{i=1}^{n}{\tilde{v}_i(x_i^{ne})}$, compared with
optimal centralized solution $\sum_{i\in \mathcal{N}}^{}{v_i(x_i^*)}$
for diffrent $L=\{0,0.1,\dots,1\}$. Figure~\ref{fig:PoA} represents
the price of anarchy in competition against two parameters of price
and number of users. The PoA is so sensitive to number of users in the
competition such that increasing number of users negatively affect on
PoA. Moreover, if the selfish behavior of users in competition do not
control by pricing, inefficiency increase and consequently the PoA
decrease. Note that the price of anarchy is independent of the
competition topology~\cite{roughgarden2003price}.
%\begin{figure}[H]
%  \centering
%  \subfloat[]{\includegraphics{contour_Csmall_tau.pdf} \label{fig:contour_Csmall_tau}}
%  \subfloat[]{\includegraphics{v_tilde_fmincon.pdf} \label{fig:v_tilde_fmincon}}
%  \caption{Inefficiency of distributed competitive resource
%    allocation, \protect\subref{fig:contour_Csmall_tau} In two-player
%    game. Nash equilibrium, \emph{i.e.}, best response functions lines
%    intersection, compared with optimal solution
%    $ \sum_{i=1}^{n}{\tilde{v}_i(x_i^{ne})}$,
%    \protect\subref{fig:v_tilde_fmincon} In $n$-player ($n=50$) game
%    compared with optimal solution.}
%\end{figure}
\subsection{Related Work} \label{Related_Work} Both centralized and
distributed solution approaches for the generic problem of resource
allocation were studied widely in various fields of expertise and a
full review is impossible here.
In data networks, that the optimization problem called \emph{Network Utility Maximization (NUM)},
users utility functions are commonly considered to be concave, continuous
and strictly increasing functions modeling \emph{elastic
  networks}~\cite{kelly1997charging},~\cite{shenker1995fundamental},
which are more mathematically tractable~\cite{boyd2004convex}, but
limits applicability. 
Many other applications require \emph{inelastic network} that are more challenging, where non-concave or discontinuous utility functions need to be maximized. Inelastic networks studied in~\cite{lee2004nonconvexity},~\cite{fazel2005network},~\cite{hande2007distributed}
and Sigmoidal programming algorithm is proposed
in~\cite{udell2013maximizing}. In~\cite{abdel2014utility}, using
utility proportional fairness policy, both elastic and inelastic
utility functions compared.\\
%In large-scale networks, distributed
%solutions are particularly attractive where a centralized solution is
%not feasible~\cite{palomar2006tutorial}.
There is also substantial literature on AIMD, the algorithm proposed
in~\cite{chiu1989analysis} and applied experimentally
in~\cite{jacobson1988congestion}, as the most
efficient-fair rate control in Internet applications. The efficiency
and fairness of the AIMD algorithm also investigated
in~\cite{lahanas2003exploiting} and a comprehensive review of the AIMD
algorithm and its applications is collected in
\cite{corless2016aimd}. This work uses the result
of~\cite{wirth2014nonhomogeneous} that used AIMD algorithm in
stochastic framework for common goods resource allocation. \\
EV charging has been the most widely studied as an application of
distributed resource allocation. In~\cite{ardakanian2013distributed},
 proposed a distributed control algorithm that
adapts the charging rate of EVs to the available capacity of the
network ensuring that network resources are used efficiently and each
EV charger receives a fair share of these resources. 
In~\cite{studli2012flexible}, proposed a
distributed AIMD based algorithm to allocate available power among
connected EVs in order to maximize the utilization of EV owners in a
range of situations.  In~\cite{studli2012aimd}, they also used
the same formalization framework to expand the modifications of the
basic AIMD algorithm to charge EVs. In both articles they considered a
fairness policy as a constraint.  The effectiveness of AIMD at
mitigating the impact of domestic charging of EVs on low-voltage
distribution networks is investigated~\cite{liu2015enhanced}.\\
\section{Conclusions \& Future Work} \label{sec:CFW} 
Our work represents a variety of AIMD-based distributed algorithms for efficient and private resource allocations in real life applications.
We considered two type of users' utility functions based on the application, first strictly increasing concave functions to represent greediness of users and second Sigmoidal functions to describe the utility from goods that are only useful in sufficient quantities.
We introduced a stochastic AIMD algorithm to allocate subsidized goods where users have concave and nonmonotonous utility functions. We also proposed derandomized version of AIMD algorithm to allocate common goods to users with strictly increasing utility functions. 
We extended the results to propose the variant of AIMD algorithm to allocate common resource where users have Sigmoidal utility functions. The numerical simulations of EV charging are also included in order to validate the convergence of our solutions.\\
%\bibliographystyle{plain}
\bibliography{BibFile1}

%\bibliographystyle{ACM-Reference-Format}  % do not change this line!
%\bibliography{BibFile1}  % put name of your .bib file here
\end{document}